\documentstyle[12pt]{article}
\begin{document}

\title{{\bf 4d neutral dilatonic black holes and (}$4+p${\bf ) dimensional
nondilatonic black }$p${\bf -branes}}
\author{J.R. Morris \\
{\it Physics Department, Indiana University Northwest,}\\
{\it 3400 Broadway, Gary, Indiana 46408 USA}}
\date{}
\maketitle

\begin{abstract}
It is shown that, in contrast to the case of extreme 4d dilatonic black
holes, 4d neutral dilatonic black holes with horizon singularities can not
be interpreted as nonsingular nondilatonic black $p$-branes in ($4+p$)
dimensions, regardless of the number of extra dimensions $p$. That is, extra
dimensions do not remove naked singularities of 4d neutral dilatonic black
holes.

\smallskip\ 


\bigskip\ 
\end{abstract}

It has been pointed out by Gibbons, Horowitz, and Townsend\cite{GHT} that
for certain values of the dilaton coupling $\alpha =\sqrt{p/(p+2)}$, where $p
$ is an odd integer, the horizon singularities of extreme 4d dilaton black
holes\cite{GM} can be removed by extra dimensions. That is, for certain
numbers of extra dimensions ($p=odd$) the extreme 4d dilaton black hole can
be viewed from a $(4+p)$ dimensional perspective as a singularity free
nondilatonic black $p$-brane. It can be inferred that this result may not
hold for the case of nonextreme dilaton black holes. In fact, Wesson and de
Leon\cite{WdL} have investigated the neutral spherically symmetric 4d black
hole solutions for 5d Kaluza-Klein theory and have shown that the horizon
singularities of the neutral nonrotating dilatonic solutions are not removed
by the extra dimension, {\it i.e.} the 5d nondilatonic neutral black 1-brane
is not singularity free, except for the Schwarzschild case. This was done by
computing the 5d Kretschmann invariant. The purpose of this paper is to
extend this result to an arbitrary number $p$ of extra dimensions. This can
be accomplished by considering the neutral, spherically symmetric 4d dilaton
solutions (in nonisotropic coordinates) of Brans-Dicke (BD) theory that
arise from a reduction of the ($4+p$) dimensional action for pure gravity.
In nonisotropic coordinates the Schwarzschild case is easily identified,
and, without appealing to scalar no hair theorems, it can be seen that all
other cases arise from singular black $p$-branes, regardless of the value of 
$p$.

In ($4+p$) dimensions the metric is given by 
\begin{equation}
ds^2=\tilde{g}_{MN}dX^MdX^N=g_{\mu \nu }dx^\mu dx^\nu -\delta
_{mn}B^2(x)dy^mdy^n,  \label{e1}
\end{equation}

\noindent where $X^M=(x^\mu ,y^m)$, with $M$,$N=0,\cdot \cdot \cdot ,p+3$, $%
\mu $,$\nu =0,\cdot \cdot \cdot ,3$, $m$,$n=1,\cdot \cdot \cdot ,p$, and the
metric has a signature diag $g_{MN}=(+,-,\cdot \cdot \cdot ,-)$. The action
for pure gravity is 
\begin{equation}
S_{(4+p)}=\frac 1{16\pi \bar{G}}\int d^4xd^py\sqrt{\tilde{g}}\,\tilde{R},
\label{e2}
\end{equation}

\noindent where $\bar{G}$ is the gravitation constant for the ($4+p$)
dimensional theory and $\tilde{g}=|\det (\tilde{g}_{MN})|$. We define the
scalar field $\varphi =\frac 1G\left( \frac B{B_0}\right) ^p$, where $B_0$
is a constant, $g=\det g_{\mu \nu }$, $R=g^{\mu \nu }R_{\mu \nu }$ is the 4d
Ricci scalar, and the usual Newtonian gravitation constant $G$ is related to 
$\bar{G}$ by $\bar{G}=GV_p$, where the ``volume'' of the compactified
internal space $V_p=B_0^p\int d^py$ is assumed to be finite. The action then
takes the form 
\begin{equation}
S=\frac 1{16\pi }\int d^4x\sqrt{-g}\left\{ \varphi R+\frac \omega \varphi
g^{\mu \nu }\partial _\mu \varphi \partial _\nu \varphi \right\}
,\,\,\,\,\,\,\,\,\,\,\omega =-1+\frac 1p\,\,.  \label{e3}
\end{equation}

\noindent This is just the action for Brans-Dicke theory in vacuum with a BD
scalar field $\varphi $ and a fixed BD parameter $\omega =-1+\frac 1p$.
Therefore, vacuum solutions for the Brans-Dicke theory corresponding to the
particular value of BD parameter $\omega =-1+\frac 1p$ are also vacuum
solutions for ($4+p$) dimensional Einstein gravity, and these solutions can
be interpreted as neutral nondilatonic black $p$-branes in ($4+p$)
dimensions.

The static spherically symmetric BD vacuum solutions \cite{BD},\cite{Brans}
have been studied previously by Campanelli and Lousto \cite{CL}, Saa\cite
{Saa} Rama \cite{Rama}, and Kim \cite{Kim}. Campanelli and Lousto (CL) have
represented the solutions in the form 
\begin{equation}
\begin{array}{cc}
ds^2=A^{m+1}dt^2-A^{n-1}dr^2-r^2A^nd\Omega ^2, &  \\ 
\varphi =\varphi _0A^{-\frac 12(m+n)},\,\,\,\,\,A=\left( 1-\frac{2r_0}%
r\right) , & 
\end{array}
\label{e4}
\end{equation}

\noindent where $d\Omega ^2=d\theta ^2+\sin ^2\theta d\phi ^2$ and $\varphi
_0$, $r_0$, $m$, and $n$ are constants. In terms of these solutions
parametrized by $m$ and $n$, the BD parameter is given by \cite{CL} 
\begin{equation}
\omega =-2\frac{(m^2+n^2+mn+m-n)}{(m+n)^2}.  \label{e5}
\end{equation}

\noindent The Schwarzschild solution is obtained for $m=n=0$, in which case $%
\varphi $, and hence the extra dimensional scale factor $B$, is a constant.
By examining the 4d Kretschmann curvature invariant $I=R_{\mu \nu \alpha
\beta }R^{\mu \nu \alpha \beta }$ Campanelli and Lousto have concluded that
no physical singularity exists at the surface $r=2r_0$ for solutions where
either $n\leq -1$ or $m=n=0$. For nonzero values of $n$ where $n>-1$ the
surface represents a naked singularity ( in the BD frame defined by the
metric and scalar field appearing in the BD action).

However, the $n\leq -1$ solutions of BD theory are not solutions of the ($%
4+p $) dimensional pure gravity theory. To see this, we define $s=m+n$ and
use the CL value of the BD parameter in (\ref{e5}), which can then be
rewritten in the form 
\begin{equation}
s=\frac{(n-1)\pm \sqrt{(n-1)^2-2n(n-2)(2+\omega )}}{(2+\omega )}.  \label{e6}
\end{equation}

\noindent For $s$ to be real valued the quantity $f=(n-1)^2-2n(n-2)(2+\omega
)$ must be nonnegative. For $-1\leq \omega \leq 0$ we find that the
constraint $f\geq 0$ excludes all values of $n\leq -1$. Therefore the ($4+p$%
) dimensional Einstein theory does not give rise to any of the 4d BD
solutions for which $n\leq -1$. However, the Schwarzschild solution is
allowed. We will make use of these facts shortly.

To determine whether the singularities of the 4d BD solutions at $r=2r_0$
can be removed by the extra dimensions, we compute the ($4+p$) dimensional
Kretschmann scalar $\tilde{I}=\tilde{R}_{ABMN}\tilde{R}^{ABMN}$ for the
nondilatonic neutral black $p$-brane. Using the symmetry properties of the
Riemann tensor, $\tilde{R}_{ABMN}=-\tilde{R}_{BAMN}=-\tilde{R}_{ABNM}=\tilde{%
R}_{MNAB}$, we can write $\tilde{I}=I+4\tilde{R}^{\alpha b\mu n}\tilde{R}%
_{\alpha b\mu n}+\tilde{R}^{abmn}\tilde{R}_{abmn}$, where the four
dimensional Kretschmann scalar $I=R^{\alpha \beta \mu \nu }R_{\alpha \beta
\mu \nu }$ has been computed by CL and can be written in the form 
\begin{equation}
I=\frac{4r_0^2}{r^6}A^{-(2n+2)}\left\{ \left( \frac{r_0}r\right)
^2J_1(m,n)+4\left( \frac{r_0}r\right) J_2(m,n)+6J_3(m,n)\right\} ,
\label{e6a}
\end{equation}

\noindent where the $J_i(m,n)$ are polynomial functions of $m$ and $n$ and
are given in \cite{CL}. A calculation for $\tilde{I}$ then leads to the
result 
\begin{equation}
\tilde{I}=\frac{4r_0^2}{r^6}A^{-(2n+2)}{\cal B}(r;m,n),  \label{e7}
\end{equation}

\noindent where 
\begin{equation}
{\cal B}(r;m,n)=\left\{ \left( \frac{2r_0}r\right) ^2K_1(m,n)+\left( \frac{%
2r_0}r\right) K_2(m,n)+K_3(m,n)\right\} ,  \label{e8}
\end{equation}

\noindent with 
\begin{equation}
\begin{array}{ll}
K_1(m,n) & =4pq^2\{-4n+4q+4 \\ 
& +\{(q+n)^2-(q+n)(n-1)+\left( \frac{p-1}2\right) q^2 \\ 
& +\frac 14[(m+1)^2+(n-1)^2+2n^2]\}\}+\frac 14J_1(m,n), \\ 
K_2(m,n) & =4pq^2(4n+4q-10)+2J_2(m,n), \\ 
K_3(m,n) & =24pq^2+6J_3(m,n),
\end{array}
\label{e9}
\end{equation}

\noindent and we have defined $q=\left( \frac{m+n}{2p}\right) $.

In order for $\tilde{I}$ to be finite on the surface $r=2r_0$, we must have
that either (1) $n\leq -1$ or (2) $n>-1$ and ${\cal B}(2r_0;m,n)=0$. As
pointed out previously, the solutions satisfying the first condition are
inaccessible to the ($4+p$) dimensional Einstein theory, so that we must
focus on solutions which satisfy the second condition, which is satisfied
for ${\cal B}(r;m,n)\sim A^2$ with $n=0$. This is equivalent to the
conditions 
\begin{equation}
K_2(m,0)=-2K_1(m,0),\,\,\,\,\,\,\,\,\,\,K_3(m,0)=K_1(m,0).  \label{e10}
\end{equation}

\noindent Using the $K_i$ in (\ref{e9}), we find that the only solution
satisfying these conditions is the one for which $m=n=0$, i.e. the
Schwarzschild solution. Therefore, except for the Schwarzschild case, the
neutral nonrotating solutions of the 4d dilatonic BD theory cannot be viewed
as nonsingular nondilatonic $p$-branes in ($4+p$) dimensions. Furthermore,
the only nondilatonic black $p$-brane solutions for which the extra
dimensions become visible\cite{DO} near $r=2r_0$ are those possessing a
singularity at $r=2r_0$.

\smallskip\ 

\newpage\

\end{document}